\documentclass[10pt,final,journal,letterpaper,twoside,twocolumn]{IEEEtran}
\usepackage{cite}
\usepackage[cmex10]{amsmath}
\usepackage{graphicx,epsfig,amssymb}
\usepackage{anysize}
\marginsize{0.8in}{0.8in}{0.25in}{0.25in} 

\begin{document}

\title{Energy Efficiency Evaluation of Cellular Networks Based on Spatial Distributions of Traffic Load and Power Consumption}
\author{Lin~Xiang,
        Xiaohu~Ge,~\IEEEmembership{Senior~Member,~IEEE,}
        Cheng-Xiang~Wang,~\IEEEmembership{Senior~Member,~IEEE,}
        Frank~Y.~Li,~\IEEEmembership{Senior~Member,~IEEE,}
        and Frank Reichert

\thanks{\scriptsize{Manuscript received December 4, 2011; revised May 22, 2012 and October 7, 2012; accepted November 2, 2012. The associate editor coordinating the review of this paper and approving it for publication was Dr. L. Deneire.}}
\thanks{\scriptsize{L. Xiang, X. Ge (Corresponding author) are with the Department of Electronics and Information Engineering, Huazhong University of Science and Technology, Wuhan 430074, Hubei, China. email: xianglin@smail.hust.edu.cn, xhge@mail.hust.edu.cn. }}
\thanks{\scriptsize{C.-X. Wang is with School of Information Science and Engineering, Shandong University, Jinan 250100, Shandong, China and Joint Research Institute for Signal and Image Processing, School of Engineering \& Physical Sciences, Heriot-Watt University, Edinburgh, EH14 4AS, UK. email: cheng-xiang.wang@hw.ac.uk.}}
\thanks{\scriptsize{F.Y. Li and F. Reichert are with the Department of Information and Communication Technology, University of Agder (UiA), Grimstad, Norway. email: \{frank.li, frank.reichert\}@uia.no.}}
\thanks{\scriptsize{L. Xiang, X. Ge and C.-X. Wang would like to acknowledge the support from the RCUK for the UK-China Science Bridges Project: R\&D on (B)4G Wireless Mobile Communications. L. Xiang, X. Ge, F.Y. Li and F. Reichert acknowledge the support from the EU FP7-PEOPLE-IRSES program, project acronym S2EuNet (Grant no.: 247083). L. Xiang and X. Ge also acknowledge the support from the National Natural Science Foundation of China (NSFC) (Grant No.: 61210002 and 61271224), National 863 High Technology Program of China (Grant No.: 2009AA01Z239) and the Ministry of Science and Technology (MOST), China, International Science and Technology Collaboration Program (Grant No.: 0903), Hubei Provincial Science and
Technology Department (Grant No.: 2011BFA004). C.-X. Wang acknowledges the support from the Scottish Funding Council for the Joint Research Institute in Signal and Image Processing with the University of Edinburgh, which is a part of the Edinburgh Research Partnership in Engineering and Mathematics (ERPem), and the Opening Project of the Key Laboratory of Cognitive Radio and Information Processing (Guilin University of Electronic Technology), Ministry of Education (No.: 2011KF01).}}
\thanks{\scriptsize{Digital Object Identifier XXXX}}
}
\maketitle

\markboth{IEEE Trans. on WIRELESS COMMUNICATIONS, Vol. XX, No. Y, Month 2012} {Xiang \MakeLowercase{\textit{et al.}}: Energy Efficiency Evaluation Based on Spatial Distributions of Traffic Load and Power Consumption}%

\begin{abstract}
Energy efficiency has gained its significance when service providers' operational costs burden with the rapidly growing data traffic demand in cellular networks. In this paper, we propose an energy efficiency model for Poisson-Voronoi tessellation (PVT) cellular networks considering spatial distributions of traffic load and power consumption. The spatial distributions of traffic load and power consumption are derived for a typical PVT cell, and can be directly extended to the whole PVT cellular network based on the Palm theory. Furthermore, the energy efficiency of PVT cellular networks is evaluated by taking into account traffic load characteristics, wireless channel effects and interference. 
Both numerical and Monte Carlo simulations are conducted to evaluate the performance of the energy efficiency model in PVT cellular networks. These simulation results demonstrate that there exist maximal limits for energy efficiency in PVT cellular networks for given wireless channel conditions and user intensity in a cell.
\end{abstract}


\section{Introduction}
\IEEEPARstart{D}{ata} traffic in cellular networks is expected to have a lasting exponential increase at least in the next five years \cite{rf12}. Currently, one of the biggest challenges is the continuous growth in energy consumption by cellular infrastructure equipment, base stations (BSs) especially, which make up to about 80\% in the total energy consumption of cellular infrastructure \cite{rf2}. To satisfy such traffic demand while still keeping energy consumption comparatively low, the need for energy efficiency improvement, i.e., the reduction of energy consumption per traffic bit has gained paramount importance for both economical and environmental advances \cite{rf47}.

Traditional traffic engineering in cellular networks was focused on traffic measurement, characterization and control, aiming to carry the largest amount of traffic load while satisfying the required quality of service (QoS) using limited radio resources, e.g., wireless channels \cite{rf1}. The new dimension of energy consumption can be similarly added into this problem, since energy is also an important type of radio resource, except that the objective now becomes to minimize energy consumption per traffic bit, for energy efficiency maximization under given QoS constraints \cite{rf49}. Typically, cellular network operators allocate their radio resources based on the \emph{worst case} principle, which is to allocate available resources based on usage requirements at peak traffic load \cite{rf50}. This leads to significant waste of resources, including energy consumption, during periods with low traffic load \cite{rf51}. To save energy, a shut-down or switch-off under-utilized BSs scheme was proposed at low traffic load conditions \cite{rf16}. A method to accommodate peak traffic load with secondary (micro-, pico- or femto-) BSs that have smaller coverages was presented in \cite{rf17}. Furthermore, combining these two methods was explored by, for example, employing the shut-down strategy to smaller secondary BSs in \cite{rf44}. An adaptive traffic coalescing (ATC) scheme was proposed to monitor and adaptively coordinate traffic packets in a mobile platform with sleep mechanisms \cite{rf52}. In our earlier work \cite{rf53}, we explored the tradeoff between the operating power and the embodied power contained in the manufacturing process of infrastructure equipments from a life-cycle perspective. From an operational perspective, we further investigated the tradeoff between transmission power and circuit power in a cellular network by optimizing the number and the location of BSs \cite{rf40}. However, most of the above traffic studies in wireless networks neglect realistic physical transmission processes, especially under wireless channel effects and interference which are significant for multi-cell wireless networks or large-scale ad hoc networks.

Improving spectral efficiency has been the primary force that drives wireless communications research over the past years. To increase spectral efficiency, adaptive modulation techniques were proposed by adapting rate \cite{rfA1} and power \cite{rfA2}, constellation size \cite{rfA3}, modulation modes \cite{rfA4} and etc. according to variations of fading channels. Besides, transmitter power control has shown significant capability to increase system capacity by suppressing co-channel interference \cite{rfA5}. The impact of power control on terminal battery duration is further explored in \cite{rfA6}. However, untill quite recently, energy efficiency in wireless communications has begun to attract lots of attention in the literature
\cite{rf18,rf19,rf20,rf21}. Energy efficiency in interfering channels was analyzed in a multi-cell wireless network \cite{rf20}. Considering frequency selective channels in orthogonal frequency division multiple access (OFDMA) systems, a new scheme which adapts the overall transmission power and its allocation according to channel states and circuit power consumption was proposed in  \cite{rf18}, in order to maximize energy efficiency. In \cite{rf19}, the link adaptation problem was formulated as a convex optimization problem and a set of link parameters were derived to maximize energy efficiency under given QoS constraints. Furthermore, a mechanism was proposed to switch between multiple input multiple output (MIMO) and single input multiple output (SIMO) modes to conserve mobile terminals' energy in the uplinks of cellular networks \cite{rf21}. However, most existing research in energy efficiency evaluation was limited at the link level of cellular networks. To analyze energy efficiency at the network level of cellular networks, a few structure characteristics of cellular networks were investigated \cite{rf11,rf3,rfA8,rf8,rf9}. Due to random motions, the locations of users cannot be modeled deterministically. Thus, statistical models, e.g., the Poisson point process \cite{rf11}, are preferable in system modeling and evaluation. The impact of spatial node distributions on the performance of wireless networks was analyzed in \cite{rf3}, in which 
the power of spatial models was illustrated for network level design.
On the other hand, the connectivity issues have been studied in ad hoc networks with nodes distributed according to the Poisson point process \cite{rfA8}.
Furthermore, at a large scale, for instance, a regional or national cellular network, a prominent characteristic of the network topology is its \emph{irregularity} in shape \cite{rf8}. Traditional cellular network studies under the simple hypothesis of hexagon topology are, however, far from enough to capture this macroscopic deployment feature. Therefore, considering that BSs and mobile stations (MSs) are generally assumed to be randomly located in an infinite plane \cite{rf9}, the network level energy efficiency needs to be further investigated in stochastic geometry cellular networks with complex wireless channel conditions.

In this paper, we propose an energy efficiency model considering traffic load and wireless channel effects in a stochastic geometry cellular network. The main contributions of this work are summarized as follows.
\begin{enumerate}
\item Compared with regular hexagonal cellular scenarios, a stochastic geometry cellular scenario with PVT cell coverages, is proposed to analyze spatial distributions of traffic load and power consumption in cellular networks. A PVT cellular scenario is able to capture the randomness of cell coverage in a realistic propagation environment in cellular networks compared with traditional regular hexagonal cellular scenarios.
\item An aggregate traffic load model used for a typical PVT cell is derived through a characteristic function, which can be extended to a spatial distribution of traffic load in PVT cellular networks.
\item Based on the signal-to-interference-ratio (SIR)-based power control process, we propose an analytical BS power consumption model for a typical PVT cell, considering path loss, log-normal shadowing and Rayleigh fading effects in wireless channels. The spatial distribution of power consumption in a PVT cellular network can be extended from this BS power consumption model.
\item Based on the spatial distributions of traffic load and power consumption in PVT cellular networks, an energy efficiency model is proposed for performance analysis.
\end{enumerate}

The rest of the paper is organized as follows. Section~\ref{sec2} describes the system model. In Section~\ref{sec3}, an aggregate traffic load model in a typical PVT cell is derived. In Section~\ref{sec4}, a BS power consumption model in a typical PVT cell with complex wireless channel effects is derived. Furthermore, based on the spatial distributions of aggregate traffic load and power consumption, an energy efficiency model for PVT cellular networks is presented and evaluated in Section~\ref{sec5}. Finally, Section~\ref{sec6} concludes this paper.

\section{System Model}
\label{sec2}
\subsection{Poisson-Voronoi Tessellation Cellular Networks}
\label{sec2-1}
Assume that both MSs\footnote{Throughout this work, MSs refer to active mobile users equipped with only one antenna.} and BSs are located randomly in the infinite plane ${\mathbb{R}^2}$. Besides, users' motions are isotropic and relatively slow, such that during an observation period, e.g., a time slot \cite{rfD1}, the relative positions of MSs and BSs are assumed to be stationary. Then, following the work in \cite{rf8,rf9}, the locations of MSs and BSs can be modeled as two independent Poisson point processes \cite{rf22}, which are denoted as
${\Pi _M} = \left\{ {{x_{Mi}}: i = 0,1,2, \cdots } \right\}$ and ${\Pi _B} = \left\{ {{y_{Bk}}: k = 0,1,2, \cdots } \right\}$,
where ${x_{Mi}}$ and ${y_{Bk}}$ are two-dimensional Cartesian coordinates, denoting the locations of the $i$th MS $M{S_i}$ and the $k$th BS $B{S_k}$, respectively. The corresponding intensities of the two Poisson point processes are ${\lambda _M}$ and ${\lambda _B}$.

\begin{figure}
\centerline{\includegraphics[width=8cm,draft=false]{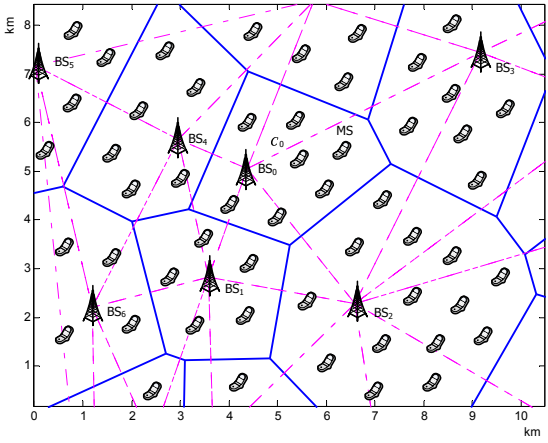}}
\caption{\small Illustration of PVT cellular structure: both BSs and MSs are randomly located; the solid lines depict cell boundaries inside which a polygon corresponds a cell coverage; and the dashed lines, which are perpendicularly bisected by corresponding cell boundaries, demonstrate how to build tessellations through the \emph{Delaunay Triangulation} method.}
\label{fig1}
\end{figure}

For a traditional cellular network, assume that an MS associates with the closest BS, which would suffer the least path loss during wireless transmission. Moreover, every cell is assumed to include only one BS and a few MSs. Then the cell boundary, which can be obtained through the \emph{Delaunay Triangulation} method by connecting the perpendicular bisector lines between each pair of BSs \cite{rf10}, splits the plane ${\mathbb{R}^2}$ into irregular polygons that correspond to different cell coverage areas. Such stochastic and irregular topology forms a so-called Poisson-Voronoi tessellation \cite{rf22}. An illustration of PVT cellular network is depicted in Fig.~\ref{fig1}, where each cell is denoted as ${{\mathcal {C}}_k}\left( {k = 0,1,2, \cdots } \right)$.

Despite of its complexity, an outstanding property of PVT cellular networks is that the geometric characteristics of any cell ${{\mathcal {C}}_k}$ with the $k$th BS centered at location ${y_{Bk}}$ (${y_{Bk}} \in {\Pi _B}$) coincide with that of a \emph{typical} PVT cell\footnote{In stochastic geometry, the \emph{typical cell} with certain geometric chatacteristic $\Xi$ is strictly defined, in ergodic means, as the empirical distribution of $\Xi$ over a square area with infinite length, which would converge to the Palm distribution of $\Xi$ and further coincide with the Palm distribution over the cell containing the origin, according to Slivnyak's theorem \cite{rf9,rf22}. More formal discussions can be found in Chapter 5 of \cite{rfC1}.} ${\mathcal{C}_0}$ where $B{S_0}$ locates at a fixed position, e.g., origin, according to the Palm theory \cite{rf9,rf22}. This feature implies that the analytical results for a typical PVT cell ${\mathcal{C}_0}$ can be extended to the whole PVT cellular network.

The propagation effects of path loss, shadowing and Rayleigh fading can be find in \cite{rfF1,rfF2}. The channel gain, defined as the ratio between the received power ${P_{rx}}$ at an MS and the corresponding transmission power ${P_{tx}}$ from its associated BS, is \cite{rfA9,rf24} 
\begin{equation}
L(r,\xi ,\zeta ) = {{{P_{rx}}} \over {{P_{tx}}}} = K{r^{ - \beta }} \cdot {e^{c\sigma \xi }} \cdot {\zeta ^2},
\label{eq2}
\tag{1}
\end{equation}
where $K$ is a constant depending on antenna gains and the term $K{r^{ - \beta }}$ models the path loss effect between the transmitting BS and the receiving MS at distance $r$ away with path loss exponent $\beta$; the term ${e^{c\sigma \xi }}$ accounts for log-normal shadowing with deviation $\sigma $, where $\xi \sim Gaussian(0,1)$ is a standard normal random variable (r.v.) and constant $c = \ln 10/10$; term ${\zeta ^2}$ is exponentially distributed with mean value 1 in the Rayleigh fading environment. Here, the path loss term $K{r^{ - \beta }}$ depends on the stochastic geometry of PVT cellular networks defined above, instead of being deterministic. For the sake of simplicity, assume that channel parameters $\beta ,\sigma $ are consistent across the whole PVT cellular network. Hereafter abbreviation ${L_{ki}}$ will be alternatively used to denote the channel gain of a link between $BS{_k}$ and $M{S_i}$ ($k,i = 0,1,2, \cdots $).

\subsection{Problem Formulation}
\label{sec2-3}
Under the assumption of PVT cellular networks, a basic framework is used to evaluate spatial distributions of traffic load and power consumption in a typical PVT cell ${\mathcal{C}_0}$. An additive functional to characterize the geometrical structure of PVT cells was proposed in \cite{rf8,rf9}, as
\begin{equation}
{\Omega_w}\mathop  = \limits^{def} \sum\limits_{{x_i} \in {\Pi _M}} {w\left( {{x_i}} \right)\mathbf{1}\left\{ {{x_i} \in {\mathcal{C}_0}} \right\}},
\label{eq3}
\tag{2}
\end{equation}
where $w(x):{\mathbb{R}^2} \to {\mathbb{R}_ + }$ is a non-negative weight function defined on each point ${x_i} \in {\Pi _M}$; $\mathbf{1}\left\{ {...} \right\}$ is an indicator function, which equals to 1 when the condition inside the bracket is satisfied and 0 otherwise; and ${\mathcal{C}_0}$ is a typical PVT cell under investigation. Based on (\ref{eq3}), we step further and study the aggregate traffic load and total BS transmission power considering spatial distributions in a typical PVT cell:

\begin{enumerate}
	\item[(1)] Aggregate traffic load ${\mathcal{T_C}_0}$ in a typical PVT cell ${\mathcal{C}_0}$
\end{enumerate}

The traffic rate, $\rho $, of the $i$th MS ${MS_i}$ at any location ${x_{Mi}}$ is denoted as $\rho (x_{Mi})$, referred as the spatial traffic intensity at ${MS_i}$. Moreover, assume that the traffic load at each MS is generated independently. Then the aggregate traffic load ${\mathcal{T_C}_0}$, which is the sum of spatial traffic intensity of all associated MSs in a typical PVT cell ${\mathcal{C}_0}$, can be defined by a revised additive functional as follows
\begin{equation}
{\mathcal{T_C}_0}\mathop  = \limits^{def} \sum\limits_{{x_{Mi}} \in {\Pi _M}} {\rho ({x_{Mi}})\mathbf{1}\left\{ {{x_{Mi}} \in {\mathcal C}_0} \right\}} .
\label{eq4}
\tag{3}
\end{equation}

\begin{enumerate}
	\item[(2)] Total BS transmission power ${\mathcal{P_C}_0}$ in a typical PVT cell ${\mathcal{C}_0}$
\end{enumerate}

By summing up the transmission power over all active downlinks in a typical PVT cell ${\mathcal{C}_0}$, the total BS transmission power can be defined as follows
\begin{equation}
    {\mathcal{P_C}_0}\mathop  = \limits^{def} \sum\limits_{{x_{Mi}} \in {\Pi _M}} {\varepsilon ({x_{Mi}})\mathbf{1}\left\{ {{x_{Mi}} \in {\mathcal{C}_0}} \right\}},
\label{eq5}
\tag{4}
\end{equation}
where $\varepsilon ({x_{Mi}})$ is the per-link transmission power from the BS to its associated MS located at ${x_{Mi}}$. According to the SIR-based power control process \cite{rf27}, the per-link transmission power $\varepsilon ({x_{Mi}})$ depends on user's traffic rate, interference and wireless channel conditions. The detailed modeling of per-link transmission power and total BS transmission power is extended in Section~\ref{sec4}.

\section{Spatial Distribution of Traffic Load in a PVT Cellular Network}
\label{sec3}

\subsection{Spatial Distribution of Traffic Load in a Typical PVT Cell}
\label{sec3-1}
Prior to characterizing the traffic load distribution, the area distribution of PVT cells needs to be investigated as a basis for later derivations. Due to the complexity of the defined PVT structure, no analytical result or even approximation about the distribution of a PVT cell area has been achieved currently. However, through large number of simulations, the empirical area distribution of PVT cells has been demonstrated to fit a Gamma distribution \cite{rf10}.

Given that the intensity of BS Poisson point process is ${\lambda _B}$, a typical PVT cell area, denoted as ${A_{\mathcal{C}0}}$, follows a Gamma distribution with the probability density function (PDF) expressed as,
\begin{equation}
{f_{A_{\mathcal{C}0}}}(x) = \frac{{{{(b{\lambda _B})}^a}}}{{\Gamma (a)}}{x^{a - 1}}\exp ( - b{\lambda _B}x),
\label{eq6a}
\tag{5}
\end{equation}
where the Gamma function $\Gamma \left(x\right) = \int_0^\infty{{t^{x - 1}}{e^{ - t}}dt}$; $a$ is the shape parameter and $b{\lambda _B}$ together is the inverse scale parameter for a Gamma distribution, respectively.

Many empirical measurement results have demonstrated that the traffic load in both wired and wireless networks, including cellular networks, is self-similar and bursty \cite{rf7}. Different from traditional traffic sources, self-similar traffic exhibits characteristics of slow-decaying tails and burstiness at different time-scales. Self-similar traffic models, e.g., Pareto distributions with infinite variance, have gained significant attention in modeling wireless networks \cite{rf56}. To evaluate the impact of self-similar traffic load on energy efficiency, the spatial traffic intensity $\rho \left( {{x_{Mi}}} \right)$ at $M{S_i}$ is assumed to be governed by a Pareto distribution with infinite variance in our study. Moreover, the spatial traffic intensity of all MSs is assumed to be independently and identically distributed (\textit{i.i.d.}). Then, a PDF of spatial traffic intensity is given by
\begin{equation}
{f_\rho }(x) = {\frac {\theta\rho_{min}^\theta} {x^{\theta + 1}}}, \qquad x \ge {\rho_{min}} > 0,
\label{eq7}
\tag{6}
\end{equation}
where $\theta  \in \left( {1,2} \right]$ reflects the \emph{heaviness} of the distribution tail. When the value of heaviness index $\theta$ is closer to one, the distribution tail of spatial traffic intensity becomes heavier. This result implies slower decaying in the tail of PDF curve and more burstiness in the defined traffic load. Parameter ${\rho _{\min }}$ represents the minimum traffic rate preserved for MS's QoS guarantee. Furthermore, the average spatial traffic intensity at MSs is 
${\mathbf{E}}\left( \rho \right) = \frac {\theta{{{\rho _{\min }}}}} { \theta - 1 }$,
where ${\mathbf{E}}( \cdot )$ denotes an expectation operation.

The spatial distribution of aggregate traffic load depends stochastically on both MS locations in space and traffic intensities that each MS point carries. To capture the randomness of both MS location and traffic intensity processes, the aggregate traffic load ${\mathcal{T}_{\mathcal{C}0}}$ in a typical cell $\mathcal{C}_0$ can be viewed as a marked Poisson point process, which is defined on the MS Poisson point process ${\Pi _M}$ \cite{rf11,rf22} with the MS points assigned \emph{i.i.d.} traffic intensity marks. The intensity measure is a random counting measure in characterizing the distribution of a spatial point process that is either homogenous or heterogeneous. For the marked traffic load point process, the intensity measure is given by
\begin{equation}
\Lambda \left( {dx,d\varphi } \right) = {\lambda _M}{A_{\mathcal{C}0}}dx \cdot {f_\rho }\left( \varphi  \right)d\varphi.
\label{eq9}
\tag{7}
\end{equation}
Then, the characteristic function of the aggregate traffic load ${\mathcal{T}_{\mathcal{C}0}}$ can be derived by conditioning on the BS Poisson point process ${\Pi _B}$\cite{rf9}
\begin{equation}
\begin{split}
 {\phi _{{\mathcal{T}_{\mathcal{C}0}}}}\left( \omega  \right) &= {{\bf{E}}_{A_{\mathcal{C}0}}}\left\{ {\exp \left[ { - {\lambda _M}A_{\mathcal{C}0}\left( {1 - {\phi _\rho }(\omega )} \right)} \right]} \right\} \\
 &= {\left[ {1 + {{{{\lambda _M}} \over {b{\lambda _B}}}} - {{{{\lambda _M}} \over {b{\lambda _B}}}}{\phi _\rho }\left( \omega  \right)} \right]^{ - a}} ,
 \end{split}
\label{eq10}
\tag{8}
\end{equation}
where ${\phi _\rho }(\omega )$ is the characteristic function of spatial traffic intensity.
%
%
Furthermore, an average aggregate traffic load expression can be derived as follows
\begin{equation}
\begin{split}
{\mathbf{E}}\left( {\mathcal{T}_{\mathcal{C}0}} \right) &= {\lambda _M}{{\mathbf{E}}_\rho}\left\{ {\exp \left[ {\int_\varphi  {{{\mathbf{1}} \left\{ {{x_{Mi}} \in {\mathcal{C}_0}} \right\} \varphi {F_\rho }\left( {d\varphi } \right)} }} \right]} \right\} \\
  &= \frac{{{\lambda _M}}}{{{\lambda _B}}}{\mathbf{E}}\left( \rho  \right) ,
\end{split}
\label{eq12}
\tag{9}
\end{equation}
where ${F_\rho }\left(  \cdot  \right)$ is the cumulative distribution function (CDF) of spatial traffic intensity.

Considering that the spatial traffic intensity is governed by a Pareto distribution, the characteristic function of ${\mathcal{T}_{\mathcal{C}0}}$ is further derived as follows
\begin{equation}
\begin{aligned}
{\phi _{{\mathcal{T}_{\mathcal{C}0}}}}\left( {j\omega } \right) &= \Bigg[ {1 + {{{{\lambda _M}} \over {b{\lambda _B}}}} - {{{{\lambda _M}} \over {b{\lambda _B}}}}\theta{{( - j{\rho_{\min}}\omega )}^\theta}}\\
& \qquad \qquad \qquad {\times \Gamma \left( -\theta, -j{\rho_{\min}}\omega \right)} \Bigg]^{-a},
\end{aligned}
\label{eq13a}
\tag{10a}
\end{equation}
with
\begin{equation}
    \Gamma ( - \theta, - j{\rho _{\min }}\omega ) = \int_{ - j{\rho _{\min }}\omega }^{+\infty } {{t^{ - \theta - 1}}{e^{ - t}}dt}
\label{13b}
\tag{10b}
\end{equation}
and the average aggregate traffic load expression ${\mathbf{E}}({\mathcal{T}_{\mathcal{C}0}})$ is simplified as
\begin{equation}
{\bf{E}}\left( {\mathcal{T}_{\mathcal{C}0}} \right) = \frac{{{\lambda _M}\theta{\rho_{\min}}}}{{{\lambda _B}(\theta - 1)}}.
\label{eq14}
\tag{11}
\end{equation}
From the characteristic function in (10), the distribution function of the aggregate traffic load can be uniquely determined by inverse Fourier transforms \cite{rf28}. Furthermore, applying the Palm theory \cite{rf9,rf22}, the aggregate traffic load model proposed in (\ref{eq10}) in a typical PVT cell can be safely extended to the whole PVT cellular network.

\subsection{Discussions on the Traffic Load Model}
\label{sec3-3}
Based on the above proposed spatial distribution of traffic load, numerical results are illustrated in this subsection. Default values for the following parameters are configured as: $a = \rm{3.61}$ and $b = \rm{3.57}$, which are based on the best-fit results \cite{rf10}; $\theta = $1.8 and ${\rho _{\min }} = \rm{10.75}$ kbps for typical wireless communication applications \cite{rf46}; and the intensity ratio of MSs to BSs is configured as ${{{\lambda _M}}}/{{{\lambda _B}}} = \rm{15}$ \cite{rf55}.

\begin{figure}
\centerline{\includegraphics[width=7.5cm,draft=false]{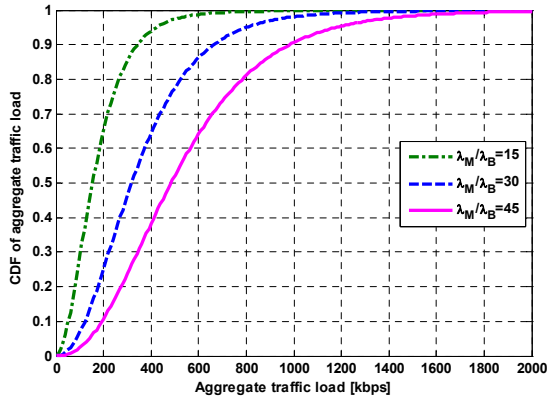}}
\caption{\small The CDF of aggregate traffic load in a typical PVT cell with different intensity ratios of MSs to BSs.}
\label{fig2}
\end{figure}

Fig.~\ref{fig2} shows the CDF of aggregate traffic load with different intensity ratios of MSs to BSs based on the Fourier transform of (\ref{eq13a}). Fig.~\ref{fig2} indicates that the probability mass shifts to the right with the increasing intensity ratio of MSs to BSs. This implies that the average aggregate traffic load increases with the increasing value of ${{{\lambda _M}}}/{{{\lambda _B}}}$, i.e., the average aggregate traffic load will increase when there are more MSs in a typical PVT cell. 

\begin{figure}
\centerline{\includegraphics[width=7.5cm,draft=false]{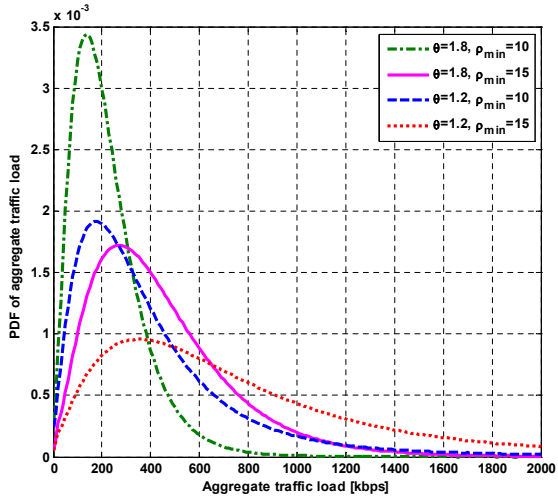}}
\caption{\small Impact of heaviness index and minimum traffic rate on the aggregate traffic load in a typical PVT cell.}
\label{fig3}
\end{figure}

Fig.~\ref{fig3} illustrates the impact of the heaviness index $\theta$ and the minimum traffic rate ${\rho _{\min }}$ on the aggregate traffic load in a typical PVT cell. When the heaviness index $\theta$ is fixed, the probability mass would shift to the right with the increasing minimum traffic rate ${\rho _{\min }}$, resulting in slight changes in the tail probability. When the minimum traffic rate ${\rho _{\min }}$ is fixed, the tail of PDF curves would become \emph{heavier}, i.e., slower decay, with the decreasing heaviness index $\theta$, which implies higher probability of large traffic load. 
The above results also demonstrate that the minimum traffic rate and heaviness index have inverse effects on the average amount of aggregate traffic load in a typical PVT cell.


\section{Spatial Distribution of Power Consumption in a PVT Cellular Network}
\label{sec4}
\subsection{Transmission Power in Wireless Downlinks}
\label{sec4-1}
\begin{figure}
\centerline{\includegraphics[width=9cm,draft=false]{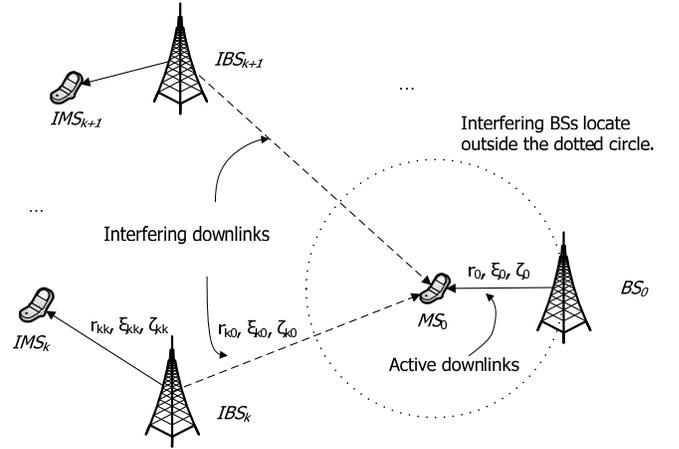}}
\caption{\small Wireless downlinks of a PVT cellular network. An example of interfering BS is illustrated at $IB{S_k}$ with detailed channel parameters.}
\label{fig5}
\end{figure}

The transmission processes of the point-to-point downlinks are power controlled by the associated BSs, in order to keep the MSs' received SIR above a given threshold.
In Fig.~\ref{fig5}, without loss of generality, a wireless downlink transmission is considered between an arbitrary MS, e.g., $M{S_0}$, and its associated BS, e.g., $B{S_0}$, inside a typical PVT cell ${{\mathcal{C}}_0}$. For $M{S_0}$, no more than one co-channel interfering MS is assumed to exist in each adjacent PVT cell. The downlinks of co-channel MSs in adjacent PVT cells contribute interference to $M{S_0}$, which are transmitted from its associated BSs. 
For distinction, a co-channel interfering MS for $M{S_0}$ is denoted as $IM{S_k}{\rm{ }}(k = 1,2,3, \cdots )$ and it is associated to a corresponding interfering BS, denoted as $IB{S_k}{\rm{ }}(k = 1,2,3, \cdots )$, in PVT cell ${{\mathcal{C}}_k}{\rm{ }}(k = 1,2,3, \cdots )$. The receiving power of $IM{S_k}$ is denoted as ${S_k}{\rm{ }}(k = 1,2,3, \cdots )$ and the receiving power of $M{S_0}$ is ${S_0}$. Moreover, all random variables ${S_k}{\rm{ }}(k = 0,1,2,\cdots )$ are assumed as i.i.d.. Then, the interference power that $M{S_0}$ receives from $IB{S_k}$ can be expressed as ${{{S_k} \cdot {L_{k0}}} \mathord{\left/
 {\vphantom {{{S_k} \cdot {L_{k0}}} {{L_{kk}}}}} \right.
 \kern-\nulldelimiterspace} {{L_{kk}}}}$ \cite{rf27}, where ${L_{kk}}$ is the channel gain of the link from $IB{S_k}$ to $IM{S_k}$ and ${L_{k0}}$ is the channel gain of the link from $IB{S_k}$ to $M{S_0}$. Neglecting the noise at MSs, the instantaneous SIR of $M{S_0}$, denoted as $\gamma $
, is given by
\begin{equation}
    \gamma  = \frac{{{S_0}}}{{{I_{{\rm{agg}}}}}} = \frac{{{S_0}}}{{\sum\limits_{k \in {\Phi _{{\rm{BS}}}}} {{S_k} \cdot \dfrac{{L({r_{k0}},{\xi _{k0}},{\zeta _{k0}})}}{{L({r_{kk}},{\xi _{kk}},{\zeta _{kk}})}} \cdot {{\bf{1}}_D}({{{L_{k0}}} \mathord{\left/ {\vphantom {{{L_{k0}}} {{L_{kk}}}}} \right. \kern-\nulldelimiterspace} {{L_{kk}}}})} }},
\label{eq19a}
\tag{12a}
\end{equation}
with
\begin{equation}
    {{\mathbf{1}}_D}({{{L_{k0}}} \mathord{\left/ {\vphantom {{{L_{k0}}} {{L_{kk}}}}} \right. \kern-\nulldelimiterspace} {{L_{kk}}}}) = {{\mathbf{1}}_D}{{\{ {r_{kk}}} \mathord{\left/ {\vphantom {{\{ {r_{kk}}} {{r_{k0}}}}} \right. \kern-\nulldelimiterspace} {{r_{k0}}}} \le 1\},
\label{eq19b}
\tag{12b}
\end{equation}
where ${I_{\rm{agg}}}$ is the aggregate interference received at $M{S_0}$, ${\Phi _{{\rm{BS}}}}$ is the index set of interfering BSs, and the indicator function ${{\mathbf{1}}_D}({{{L_{k0}}} \mathord{\left/
 {\vphantom {{{L_{k0}}} {{L_{kk}}}}} \right.
 \kern-\nulldelimiterspace} {{L_{kk}}}})$ is a constraint on MS distance distributions under the closest association rule\footnote{An MS cannot change its association BS unless it moves into another PVT cell.} in PVT cellular networks.

Without considering the additional transmission power for outage loss in downlinks, the relationship of the required receiving SIR\footnote{The SIR here approximates the signal-to-interference-noise-ratio (SINR) for cellular systems when the noise power is negligible.} $\gamma$ and the spatial traffic intensity $\rho \left( {{x_{M0}}} \right)$ at $M{S_0}$ is given by \cite{rf33}
\begin{equation}
{B_W} \cdot {\log _2}(1 + {\gamma  \mathord{\left/
 {\vphantom {\gamma  \Delta }} \right.
 \kern-\nulldelimiterspace} \Delta }) = \rho \left( {{x_{M0}}} \right) ,
\label{eq20}
\tag{13}
\end{equation}
where ${B_W}$ is the bandwidth allocated for $M{S_0}$; $\Delta $ defines the SIR gap between the channel capacity and a practical coding and modulation scheme achieving a target bit error rate at $M{S_0}$. Based on the PDF of spatial traffic intensity in (\ref{eq7}), the PDF of the receiving SIR $\gamma $ can be further derived as follows
\begin{equation}
\begin{split}
 {f_\gamma }(z) &= \frac{{\theta\rho _{\min }^\theta B_W^{ - \theta}}}{{\ln 2 \cdot (\Delta  + z)}}{\left( {{{\log }_2} \left(1 + {z \mathord{\left/
 {\vphantom {z \Delta }} \right.
 \kern-\nulldelimiterspace} \Delta } \right)} \right)^{ - \theta - 1}},\\
& \quad \quad \quad z > {z_0} = \Delta  \cdot \left({2^{{{{\rho _{\min }}}}/{B_W}}} - 1 \right) ,
\end{split}
\label{eq21}
\tag{14}
\end{equation}
where ${z_0}$ is the minimum required SIR to achieve the minimum traffic rate ${\rho _{\min }}$ for an MS.

Denote the interfering downlink set ${\Pi _{{\rm{Inf}}}}$ as the collection of co-channel active downlinks, which contains the link pairs from the interfering BS $IB{S_k}$ at location ${y_{Ik}}$ to the interfering MS $IM{S_k}$ at location ${x_{Ik}}$, i.e., ${\Pi _{{\rm{Inf}}}} = \left\{ {({x_{Ik}},{y_{Ik}}): k = 1,2,3, \cdots } \right\}$.
The interfering downlink set ${\Pi _{{\rm{Inf}}}}$ can be viewed as an independent thinning process on the BS Poission point process ${\Pi _B}$. Thus, ${\Pi _{{\rm{Inf}}}}$ is still a Poisson point process with intensity ${\lambda _{{\rm{Inf}}}}$, which generally satisfies $0 \le {\lambda _{{\rm{Inf}}}} \le {\lambda _B}$\footnote{Actually, for cellular networks with full frequency reuse, $0 \le {\lambda _{{\rm{Inf}}}} \le {\lambda _{{{B}}}}$; while for traditional cellular networks with frequency reuse ${N_f}$ (${N_f}$ is an integer and ${N_f}>1$), $0 \le {\lambda _{{\rm{Inf}}}} \le {\textstyle{{{\lambda _{{{B}}}}} \over {{N_f}}}}$. The results in this study are also applicable for the latter case.}.

Based on the stochastic geometry theory, the co-channel interference ${I_{{\rm{agg}}}}$ aggregated at $M{S_0}$ can be further expressed as follows
\begin{equation}
\begin{split}
{I_{{\rm{agg}}}} = \sum\limits_{\left({x_{Ik}}, {y_{Ik}}\right) \in {\Pi _{{\rm{Inf}}}}} &{S_k} \cdot \frac{{{{\left\| {{y_{Ik}} - {x_{Ik}}} \right\|}^\beta }}}{{{{\left\| {{y_{Ik}} - {x_{M0}}} \right\|}^\beta }}} \cdot {Q_k} \\
& \cdot {{\mathbf{1}}_{D1}}\left\{ {{x_{M0}} \in {{\mathcal C}_0}} \right\} \cdot {{\mathbf{1}}_{D2}}\left\{ {{x_{Ik}} \in {{\mathcal C}_k}} \right\},
\end{split}
\label{eq23a}
\tag{15a}
\end{equation}
with
\begin{equation}
{Q_k} = {e^{c\sigma ({\xi _{k0}} - {\xi _{kk}})}} \cdot {\frac {\zeta _{k0}^2} {\zeta _{kk}^2}} ,
\label{eq23b}
\tag{15b}
\end{equation}
where $\left\| {{y_{Ik}} - {x_{Ik}}} \right\|$ and $\left\| {{y_{Ik}} - {x_{M0}}} \right\|$ denote the Euclid distances from $IB{S_k}$ to $IM{S_k}$ and from $IB{S_k}$ to $M{S_0}$, the indicator functions ${{\mathbf{1}}_{{D_1}}}$ and ${{\mathbf{1}}_{{D_2}}}$ specify the range of distances from MSs to their associated BSs in residing PVT cells. Note that constraint (\ref{eq19b}) is met immediately.

Under the assumption that channel parameters $\beta$ and $\sigma $ are identical across the whole cellular network, ${Q_k}(k = 1,2,3, \cdots )$ are i.i.d. random variables with the following PDF
\begin{equation}
{f_{Q_k}}(x) = \frac{1}{{2\sqrt \pi  c\sigma }}\int_0^{\infty} {\frac{{\exp \left( { - {{{{t^2}} \over {4{c^2}{\sigma ^2}}}} + t} \right)}}{{{{\left( {{e^t} + x} \right)}^2}}}dt}, \quad x \ge 0 .
\label{eq24}
\tag{16}
\end{equation}
Furthermore, based on the result from the Appendix, the characteristic function of co-channel interference ${I_{{\rm{agg}}}}$ aggregated at $M{S_0}$ is derived as follows
\begin{equation}
{\phi _{{I_{{\rm{agg}}}}}}(\omega ) = \exp \left\{ { - \delta |\omega {|^{{\frac 2 \beta }}}\left[ {1 - j {\rm{sign}}(\omega )\tan \left( {{\frac \pi  \beta}} \right)} \right]} \right\},
\label{eq25a}
\tag{17a}
\end{equation}
with
\begin{equation}
\delta  = \frac{{{\lambda _{{\rm{Inf}}}}}}{{4{\lambda _B}}}\Gamma (1 - {\frac 2  \beta})\cos ({\frac \pi \beta}){\mathbf{E}}(S_k^{2/\beta }){\bf{E}}(Q_k^{2/\beta }),
\label{eq25b}
\tag{17b}
\end{equation}
\begin{equation}
{\mathbf{E}}(Q_k^{2/\beta }) = \frac{{2\pi }}{{\beta \sin (2\pi /\beta )}}\exp \left( {{\frac{4{c^2}{\sigma ^2}} {{\beta ^2}}}} \right),
\label{eq25c}
\tag{17c}
\end{equation}
where ${\text{sign}}(\omega )$ is a sign function \cite{rf54}. From (\ref{eq25a}), the aggregate interference is an $\alpha$-stable process with infinite expectation \cite{rf24,rfGe}.

Based on the results of the required SIR and interference derivations, the required receiving power can then be derived based on (\ref{eq19a}). Given that the moment of the required receiving power is ${\mathbf{E}}(S_k^{2/\beta })$, then, the characteristic function of ${S_k}(k = 0,1,2, \cdots )$ is derived \cite{rf28}
\begin{equation}
 \begin{split}
 {\phi _{{S_k}}}(\omega ) &= \int _x {{\phi _{{I_{\rm{agg}}}}}(\omega x) {f_\gamma }(x)dx}  \\
  &= \int _0^{\infty} {\exp {\left\{ \begin{array}{l}
  - \delta |\omega {|^{\frac{2}{\beta }}}{x^{\frac{2}{\beta }}} \times \\
 {\quad} \left[ {1 - j{\rm{sign}}(\omega )\tan \frac{\pi }{\beta }} \right] \\
 \end{array} \right\}} {f_\gamma}(x)dx}  .
 \end{split}
\label{eq26}
\tag{18}
\end{equation}

Furthermore, based on (\ref{eq26}), the corresponding required transmission power for the downlink from $B{S_0}$ to $M{S_0}$ is derived as follows
\begin{equation}
\begin{split}
\varepsilon \left(x_{M0}\right) &= \frac{{{S_0}}}{{L\left(\left\| {x_{M0}-y_{B0}} \right\|,{\xi_0},{\zeta_0}\right)}} \\ &= \frac{{{{\left\| {x_{M0}-y_{B0}} \right\|}^\beta } \cdot {e^{ - c\sigma {\xi_0}}}}}{{K\zeta_0^2}}{S_0} .
\end{split}
\label{eq27}
\tag{19}
\end{equation}

\subsection{Total BS Transmission Power}
\label{sec4-2}
 It is assumed that all wireless downlinks are assigned appropriate transmission power to transmit data with required traffic rate. Denote r.v.s $U = {S_0} \cdot V $ and $V = {{{e^{ - c\sigma {\xi_0} }}} \mathord{\left/
 {\vphantom {{{e^{ - c\sigma {\xi_0} }}} {{\zeta_0^2}}}} \right.
 \kern-\nulldelimiterspace} {{\zeta_0^2}}}$. By substituting (\ref{eq27}) into (\ref{eq5}), the required total BS transmission power in a typical PVT cell ${{\mathcal C}_0}$ is given by
\begin{equation}
{\mathcal{P}_{\mathcal{C}0\_\rm{req}}} = \sum\limits_{{x_{Mj}} \in {\Pi _{M}}} {{\frac{{{\left\| {{x_{Mj}} - {y_{B0}}} \right\|}^\beta }} K}U \cdot {\mathbf{1}}\left\{ {{x_{Mj}} \in {\mathcal{C}_0}} \right\}} ,
\label{eq28a}
\tag{20}
\end{equation}
where we similarly have ${\mathbf{E}}\left( {{\mathbf{1}}\left\{ {{x_{Mj}} \in {{\mathcal C}_0}} \right\}} \right) = {e^{ - \pi {\lambda _B}{{\left\| {{x_{Mj}}} \right\|}^2}}}$ and the PDF of V, ${f_V}(x)$, as follows
\begin{equation}
{f_V}(x) = \frac{1}{{\sqrt {2\pi } c\sigma {x^2}}}\int_{ - \infty }^\infty  {\exp \left( { - \frac{{{t^2}}}{{2{c^2}{\sigma ^2}}} + t - \frac{{{e^t}}}{x}} \right)dt} {\rm{ }}.
\label{eq52}
\tag{21}
\end{equation}

Furthermore, the required total BS transmission power can be also formulated by a marked Poisson point process based on MS Poisson point process with marks capturing the required transmission power of downlinks to each MS point. Different from the spatial traffic load process, here the probabilistic marks of transmission power depend on the distances between MS and BS points instead of being \emph{i.i.d.} random variables. Moreover, the spatial geometry of PVT cellular networks has greater impact on the spatial power consumption process due to its effects on downlink distance distributions, co-channel interference, and per-downlink transmission power, as illustrated in (\ref{eq27}). Following the derivations in the previous section, the intensity measure of the marked transmission power process can be similarly given as
\begin{equation}
\Lambda (dx,du) = {\lambda _M}dx \cdot {f_U}(u)du ,
\label{eq29}
\tag{22}
\end{equation}
where ${f_U}(u)$ is the PDF of $U$. Applying the Campbell theorem for marked Poisson point processes \cite{rf22}, the characteristic function of the required total BS transmission power is derived as (\ref{eq31aa}), where ${\phi _U}( \cdot )$ is the characteristic function of $U$, and we have  ${\phi _U}(\omega ) = \int _x {{\phi _{{S_0}}}(\omega x) {f_V}(x)dx}$ \cite{rf28}. By substituting $\phi _{{S_k}}(\omega )$ into (\ref{eq26}), we can further express $\phi_{{\mathcal{P}_{\mathcal{C}0\_\rm{req}}}}(\omega)$ as (\ref{eq31a}) with (\ref{eq31b}).
\begin{figure*}[!t]
\normalsize
\begin{equation}
\begin{split}
{\phi _{{\mathcal{P}_{\mathcal{C}0\_\rm{req}}}}}(\omega ) &= {\mathbf{E}}\left\{ {\exp \left[ {2\pi {\lambda _M}\int _x \! {\int _u {\left( {{e^{{{\frac{j\omega {x^\beta }u}  K}}}} - 1} \right) {f_U}(u)du \cdot {\mathbf{1}}\left\{ {x \in {\mathcal{C}_0}} \right\} xdx} } } \right]} \right\} \\
&= \exp \left[ { - 2\pi {\lambda _M}\int _0^{\infty} {\left( {1 - {\phi _U} \left({\textstyle{\frac{\omega {x^\beta }} K}} \right)} \right) {e^{ - \pi {\lambda _B}{x^2}}} xdx} } \right] ,
\end{split}
\label{eq31aa}
\tag{23a}
\end{equation}
\begin{equation}
\begin{split}
{\phi _{{\mathcal{P}_{\mathcal{C}0\_\rm{req}}}}}(\omega ) &= \exp \left\{ { - {\lambda _M}\iiint\limits_{r,x,y} {\left[ {1 - \exp \left( { - G(\tfrac{\omega }
{K}){r^2}{x^{\tfrac{2}
{\beta }}}{y^{\tfrac{2}
{\beta }}}} \right)} \right]2\pi r{e^{ - \pi {\lambda _B}{r^2}}} {f_V}(x){f_\gamma }(y)drdxdy}} \right\} \\
&= \exp \left\{ { - \frac{{{\lambda _M}}}{{{\lambda _B}}}\left[ {1 - {{\mathbf{E}}_{\gamma,V}}\left( {\frac{{\pi {\lambda _B}}}{{G({\textstyle{\omega  \over K}}){V^{\frac 2 \beta}}{\gamma ^{\frac 2 \beta}} + \pi {\lambda _B}}}} \right)} \right]} \right\} ,
\end{split}
\label{eq31a}
\tag{23b}
\end{equation}
\begin{equation}
G(\omega ) = \delta |\omega {|^{\frac 2 \beta}}\left[ {1 - j {\rm{sign}}(\omega ) \tan {{\pi  \over \beta }}} \right] .
\label{eq31b}
\tag{23c}
\end{equation}
\end{figure*}

However, numerical results show that the expectation value of the required total BS transmission power approximates infinity in the case of perfect power control. It is caused by a result that an infinite demand for transmission power is required to maintain a given SIR level when the expectation of aggregate interference approaches infinity.

Considering that perfect power control is an ideal case, a maximal BS transmission power threshold ${P_{\max }}$ is configured in the following. In this case, a portion of wireless downlinks may be interrupted when the required total BS transmission power exceeds ${P_{\max }}$. Therefore, by \emph{truncating} ${\mathcal{P}_{\mathcal{C}0\_\rm{req}}}$ in the interval $\left( {0,{P_{\max }}} \right]$, the PDF of realistic total BS transmission power in a typical PVT cell ${{\mathcal C}_0}$ can be derived as
\begin{equation}
{f_{{\mathcal{P}_{\mathcal{C}0\_\rm{real}}}}}(x) = \left\{ {\begin{array}{*{20}{c}}
   {{\dfrac {{\mathcal{P}_{\mathcal{C}0\_\rm{req}}}(x)} {{F_{{\mathcal{P}_{\mathcal{C}0\_\rm{req}}}}}({P_{\max }})}},} \qquad {x \le {P_{\max }}} ;  \\
   { {0,}    \quad\quad\quad\quad\quad\qquad {x > {P_{\max }}}    ;}  \\
\end{array}} \right.
\label{eq33}
\tag{24}
\end{equation}
where ${F_{{\mathcal{P}_{\mathcal{C}0\_\rm{req}}}}(\cdot)}$ is the CDF of ${\mathcal{P}_{\mathcal{C}0\_\rm{req}}}$.
\subsection{BS Power Consumption Model}
\label{sec4-3}
The BS power consumption can be decomposed into a fixed power consumption part and a dynamic power consumption part respectively \cite{rf13}. The fixed power consumption, e.g., the circuit power consumption, is the baseline power consumed during processes such as signal processing, site cooling, power supply and battery backup. The circuit power consumption usually depends on the hardware and software configurations of BSs and is independent of the spatial distribution of traffic load. The dynamic power consumption, depending on the spatial distribution of traffic load however, accounts for the transmission power consumed in the radio frequency (RF) transmission circuits including power dissipation such as heat.

Based on the decomposition of BS power consumption, the probability distribution of total BS power consumption can be derived from the transmission power distributions if the efficiency of RF transmission circuits is properly modeled.  However, only the average RF efficiency is normally available in practical applications. Thus, a linear average BS power consumption model is simply built as follows
\begin{equation}
\begin{split}
{\bf{E}}({P_{BS}}) &= \frac{{{\bf{E}}\left({\mathcal{P}_{\mathcal{C}0\_\rm{real}}}\right)}}{{{\eta _{{\rm{RF}}}}}} + {P_{{\rm{Circuit}}}} \\ &= \dfrac{{\int_0^{{P_{\max }}} {x{f_{{\mathcal{P}_{\mathcal{C}0\_\rm{req}}}}}(x)dx} }}{{{\eta _{{\rm{RF}}}} \cdot \int_0^{{P_{\max }}} {{f_{{\mathcal{P}_{\mathcal{C}0\_\rm{req}}}}}(x)dx} }} + {P_{{\rm{Circuit}}}} ,
\end{split}
\label{eq35}
\tag{25}
\end{equation}
where ${\eta _{{\text{RF}}}}$ is the average efficiency of RF transmission circuits and the circuit power ${P_{{\rm{Circuit}}}}$ is fixed as a constant. Applying the Palm theory \cite{rf9,rf22}, (\ref{eq35}) can be directly extended to a spatial distribution of power consumption in PVT cellular networks.

\subsection{Discussions on BS Power Consumption}
\label{sec4-4}
Based on the proposed model, we illustrate BS power consumption numerically in this subsection. The parameters for a PVT cellular network shown in Fig.~\ref{fig5} are configured as follows: the moment of receiving power is ${\mathbf{E}}(S_k^{2/\beta }) = {10^{-10}}$\:W (or -70\:dBm), which corresponds to a receiving power level of MSs in the order of ${10^{-15}}$\:W (or -120\:dBm) \cite{rf46}; the BS intensity is configured as ${\lambda _B} = {1 \mathord{\left/ {\vphantom {1 {(\pi  * {{800}^2})}}} \right.  \kern-\nulldelimiterspace} {(\pi  * {{800}^2})}}{\text{ }}{{\text{m}}^{{\text{-2}}}}$; the intensity ratio of MSs to BSs is configured as ${\lambda _M}/{\lambda _B=}$ 30 and the intensity ratio of interfering links to BSs is configured as ${\lambda _{{\rm{Inf}}}}/{\lambda _B} = 0.9$; the path-loss exponent $\beta =$ 3.5, shadowing deviation $\sigma=$ 6 and $K = -31.54{\rm{\:dB}}$ in an urban micro-cell environment \cite{rf42}; the heaviness index $\theta = 1.8$ and the SIR gap $\Delta  = 8.6{\text{\:dB}}$ are configured for a quadrature amplitude modulation (QAM) system \cite{rf33}; the channel bandwidth ${B_W}$ is normalized, then the normalized minimum rate is ${\rho _{\min }} = 2{\text{\:bits/s/Hz}}$, which corresponds to an equivalent average spectral efficiency 4.5\:bits/s/Hz for typical cellular networks \cite{rf55}.

\begin{figure}
\centerline{\includegraphics[width=7.4cm,draft=false]{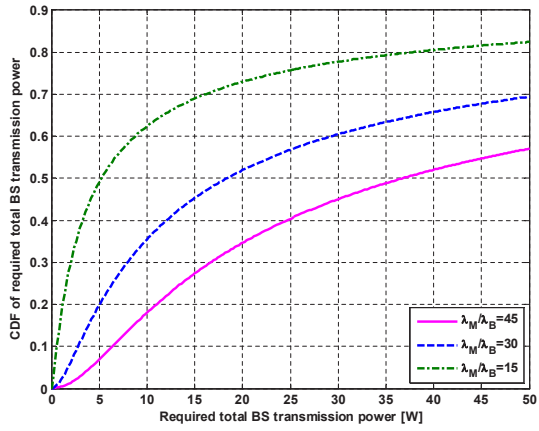}}
\caption{\small The CDF of required total BS transmission power with different intensity ratios of MSs to BSs.}
\label{fig7}
\end{figure}

Based on the inverse Fourier transform of (\ref{eq31a}), the CDF of required total BS transmission power ${\mathcal{P}_{\mathcal{C}0\_\rm{req}}}$ under different intensity ratios of MSs to BSs is illustrated in Fig.~\ref{fig7}. For a micro BS, the values of the required transmission power vary from several Watts to tens of Watts in this numerical result \cite{rf13}. With the increase of the intensity ratio of MSs to BSs, the average number of MSs in a PVT cell increases, leading to the increased total BS transmission power. Different from the CDF of aggregate traffic load, these curves exhibit a heavy tail feature, indicating the high probability of large required total BS transmission power.

\begin{figure}
\centerline{\includegraphics[width=7cm,draft=false]{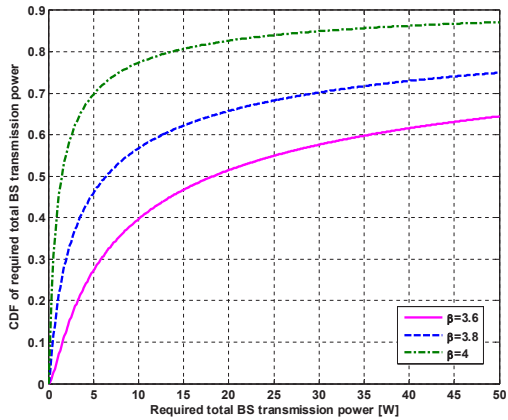}}
\caption{\small The CDF of required total BS transmission power with different path loss exponents.}
\label{fig72}
\end{figure}

Fig.~\ref{fig72} shows the CDF of required total BS transmission power with different path loss exponents. The increased path loss exponent would shift the probability mass to the left, indicating that lower transmission power is needed with larger path-loss exponent on average due to greater attenuation of interference power. The heavy tail is also easily observed. From Fig. 5 and Fig. 6, the required total BS transmission power exhibits a significant degree of bustiness, indicating higher demand for large transmission power to support self-similar traffic load.

\begin{figure}
\centerline{\includegraphics[width=7.1cm,draft=false]{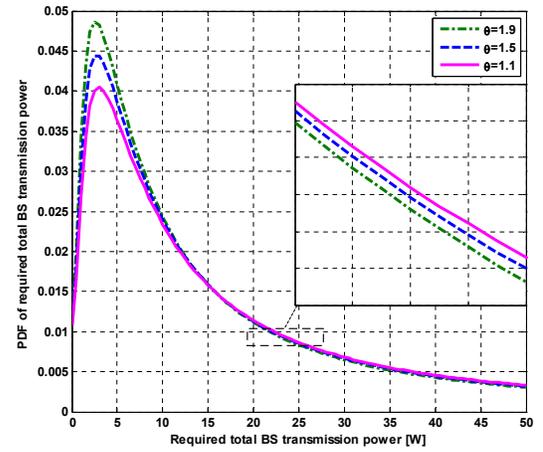}}
\caption{\small The PDF of required total BS transmission power with different heaviness indices.}
\label{fig6}
\end{figure}

In Fig.~\ref{fig6}, the PDF of required total BS transmission power ${\mathcal{P}_{\mathcal{C}0\_\rm{req}}}$ is depicted with different heaviness indices $\theta$. With the decrease of heaviness index, indicating that the traffic at MSs is more bursty, the probability mass of the required total BS transmission power remains rather stable except that the increasingly \emph{heavier} tail decays slower.

\begin{figure}
\centerline{\includegraphics[width=7.3cm,draft=false]{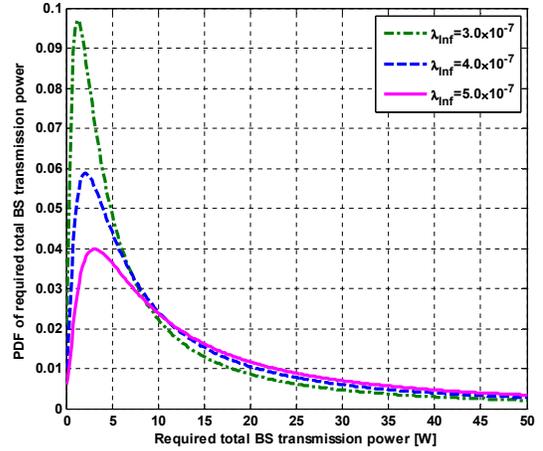}}
\caption{\small The PDF of required total BS transmission power with different interfering link intensities.}
\label{fig8}
\end{figure}

Finally, Fig.~\ref{fig8} evaluates the required total BS transmission power ${\mathcal{P}_{\mathcal{C}0\_\rm{req}}}$ with different interfering link intensities ${\lambda _{\rm{Inf}}}$. With the increase of the interfering link intensity ${\lambda _{\rm{Inf}}}$, the PDF of the required total BS transmission power would shift its probability mass to the right. This result indicates that, to guarantee the traffic rate for a specified MS, more transmission power is required to compensate for the corresponding SIR degradation with increased interference from adjacent PVT cells.

\section{Energy Efficiency of PVT Cellular Networks}
\label{sec5}
\subsection{Energy Efficiency Model}
\label{sec5-1}

\begin{figure*}[!t]
\begin{equation}
{\eta _{{\rm{EE}}}} = \frac{{{\lambda _M}\theta{{{\rho _{\min }}}}}}{{{\lambda _B}(\theta - 1)}} \cdot \frac{{{{\left( {\int_0^{{P_{\max }}} {{f_{{\mathcal{P}_{\mathcal{C}0\_\rm{req}}}}}(x)dx} } \right)}^2}}}{{{\textstyle{1 \over {{\eta _{{\rm{RF}}}}}}}\int_0^{{P_{\max }}} {x{f_{{\mathcal{P}_{\mathcal{C}0\_\rm{req}}}}}(x)dx}  + {P_{{\rm{Circuit}}}} \cdot \int_0^{{P_{\max }}} {{f_{{\mathcal{P}_{\mathcal{C}0\_\rm{req}}}}}(x)dx} }},
\label{eq38}
\tag{28}
\end{equation}
\end{figure*}

Based on the obtained spatial distributions of traffic load and power consumption in PVT cellular networks, we further investigate the energy efficiency of PVT cellular networks, which is represented by the BS energy efficiency in a typical PVT cell according to the Palm theory \cite{rf9,rf22}. The utility function of energy efficiency ${\eta _{{\text{EE}}}}$ is defined as the ratio of the effective average traffic load\footnote{Here ``effective" means the realistic traffic load eliminating the traffic load in outage, by multiplying the average aggregate traffic load $\mathbf{E}({{\mathcal{T}}_{\mathcal{C}0}})$ with the probability $1 - {p_{{\rm{out}}}}$.} over the average total power $\mathbf{E}({{P}_{BS}})$ consumed at a BS in a typical PVT cell
\begin{equation}
{\eta _{{\rm{EE}}}} = \frac{{{\bf{E}}({\mathcal{T}_{\mathcal{C}0}}) \cdot \left( {1 - {p_{{\rm{out}}}}} \right)}}{{{\bf{E}}({P_{BS}})}},
\label{eq36}
\tag{26}
\end{equation}
where the outage probability\footnote{Outage probability can be also similarly defined as the probability that the bit (symbol) error probability exceeds a maximal threshold, to measure the bit (symbol) error outage in wireless fading channels \cite{rfB1,rfB2,rfB3}.}, defined as the probability that the required transmission power fails to be satisfied under the maximal transmission power constraint due to variations of wireless channel status and effects of interference, is given by,
\begin{equation}
\begin{split}
{p_{{\rm{out}}}} &= \Pr \left\{ {{\mathcal{P}_{\mathcal{C}0\_\rm{req}}} > {P_{\max }}} \right\} \\
&= 1 - {F_{{\mathcal{P}_{\mathcal{C}0\_\rm{req}}}}}\left({P_{\max }}\right).
\end{split}
\label{eq37}
\tag{27}
\end{equation}
Based on (\ref{eq14}) and (\ref{eq35}), the energy efficiency model can be further derived as shown in (\ref{eq38}),
where the PDF of the required BS transmission power ${f_{{\mathcal{P}_{\mathcal{C}0\_\rm{req}}}}}(x)$ can be numerically calculated by its characteristic function in (\ref{eq31a}).

\subsection{Simulation Results and Discussions}
\label{sec5-2}
Based on the proposed energy efficiency model, the effect of traffic load and wireless channel parameters on the energy efficiency of PVT cellular networks is investigated in details. Furthermore, the analysis results are validated through Monte-Carlo (MC) simulations in a PVT cellular network with radius $R=10\:km$.
Different from the statistical treatment of transmission power in analytical models (e.g., in (\ref{eq33})), the maximal available transmission power ${P_{\max }} = {\text{40\:W}}$ is defined in the MC simulations and MSs are randomly selected in outage when the total transmission power reaches the maximum.
In the following, some default parameters are specified: the moment of receiving power is configured as ${\mathbf{E}}(S_k^{2/\beta }) = {10^{-10}}$\:W (or -70\:dBm); the BS intensity is given by ${\lambda _B} = {1 / {(\pi \times {{800}^2})}}{\text{ }}{{\text{m}}^{{\text{-2}}}}$, while the intensity ratio of interfering links to BSs is configured as ${\lambda _{\rm{Inf}}}/{\lambda _B} = 0.8$; the path-loss exponent $\beta = 3.8$, shadowing deviation $\sigma = 6$ and $K = {\rm{-31.54\:dB}}$ corresponding to an urban micro-cell environment; the heaviness index is $\theta = \rm{1.8}$ and the SIR gap is $\Delta = {\text{8.6\:dB}}$ for a QAM system; the normalized minimum rate is ${{\rho _{\min }}} = {\text{2\:bits/s/Hz}}$, which corresponds to an approximate average spectral efficiency level of 4.5\:bits/s/Hz for typical cellular networks; the maximal transmission power is ${P_{\max }} = {\text{40\:W}}$ \cite{rf13}; the average efficiency of RF circuit for a micro BS is configured as ${\eta _{{\text{RF}}}} = 0.047$ and the fixed BS power consumption is set as ${P_{{\text{const}}}} = {\text{354.4\:W}}$ \cite{rf34}.

\begin{figure}
\centerline{\includegraphics[width=7.2cm,draft=false]{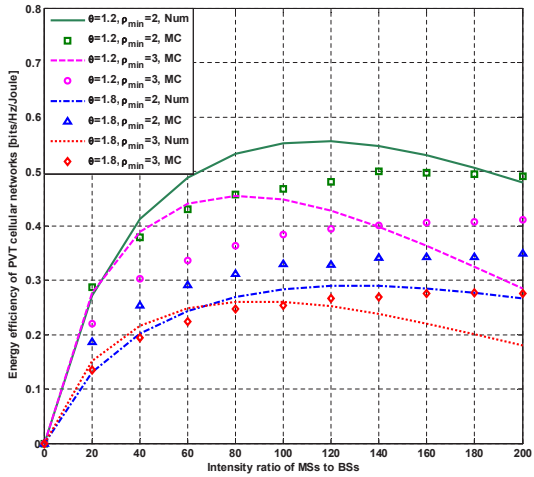}}
\caption{\small Energy efficiency of PVT cellular networks with respect to the intensity ratio of MSs to BSs considering the heaviness index and the minimum traffic rate.}
\label{fig9}
\end{figure}

Fig.~\ref{fig9} illustrates the energy efficiency of PVT cellular networks with respect to the heaviness index $\theta$ and the minimum traffic rate ${\rho _{\min }}$ at MSs, in which ``Num" labels the numerical results and ``MC" represents the MC simulation results. First, we fix the values of minimum traffic rate ${\rho _{\min }}$ at 2 or 3\:bits/s/Hz, and analyze the impact of the heaviness index $\theta$ on the energy efficiency of PVT cellular networks. Both numerical and MC simulation results consistently demonstrate that the energy efficiency of PVT cellular networks is increased when the heaviness index $\theta$ is decreased from 1.8 to 1.2. Secondly, we fix the values of the heaviness index $\theta$ at 1.2 or 1.8, and analyze the impact of the minimum traffic rate ${\rho _{\min }}$ on the energy efficiency of PVT cellular networks. The MC simulation results show that the energy efficiency of PVT cellular networks monotonically increases when the minimum traffic rate increases from 2 to 3\:bits/s/Hz. However, the numerical results show that when the heaviness index $\theta$ is fixed at either 1.2 or 1.8, there exist turning points for the intensity ratios of MSs to BSs (the turning points are 25 for $\theta =1.2$ and 68 for $\theta =1.8$). Below the turning point, the energy efficiency of PVT cellular networks increases and above the turning point the energy efficiency decreases, as the minimum traffic rate ${\rho _{\min }}$ is increased from 2 to 3 bits/s/Hz.

The numerical results in Fig.~\ref{fig9} illustrate that there exists a maximal value for the energy efficiency under each parameter configuration. The maximal energy efficiency values are 0.55, 0.45, 0.29 and 0.26, corresponding to the intensity ratios of MSs to BSs of 110, 80, 130 and 90, respectively. This result can be explained from (\ref{eq38}). When the intensity ratio of MSs to BSs is low, indicating a few MSs in a typical PVT cell, the increase of the intensity ratio of MSs to BSs leads to a moderate increase in total BS power consumption including mainly \emph{fixed} BS power consumption and a small portion of \emph{dynamic} BS power consumption. In this case, the energy efficiency of PVT cellular networks is increased. However, when the intensity ratio of MSs to BSs in a PVT typical cell exceeds a given threshold, a high aggregate traffic load resulted from a large number of MSs will significantly increase the total BS power consumption including mainly \emph{dynamic} BS power consumption and a small portion of \emph{fixed} BS power consumption. In this case, the energy efficiency of PVT cellular networks is decreased.
Such an energy efficiency pattern with respect to traffic load was also observed in wireless LANs \cite{rfE1}, where random medium accessing protocols are deployed. However, the MC simulation results in Fig. 9 demonstrate that the energy efficiency turns into saturation after having reached the maximal threshold.

The different trends between numerical and MC simulations after having reached the maximal threshold are caused by different outage control methods. In the numerical analysis, if the required transmission power exceeds the maximal transmission power ${{P}_{\max }}$ with a large number of users, MSs' traffic load requests are interrupted with the probability calculated by (\ref{eq37}). This will cause the energy efficiency to decrease after having reached the maximal threshold. In the MC simulation, for the same scenario, an MS' traffic load request would be interrupted instantly when the transmission power is unavailable. This will achieve better utilization, on the average, of the BS transmission power and turns the energy efficiency into saturation after having reached the maximal threshold.

\begin{figure}
\centerline{\includegraphics[width=7.2cm,draft=false]{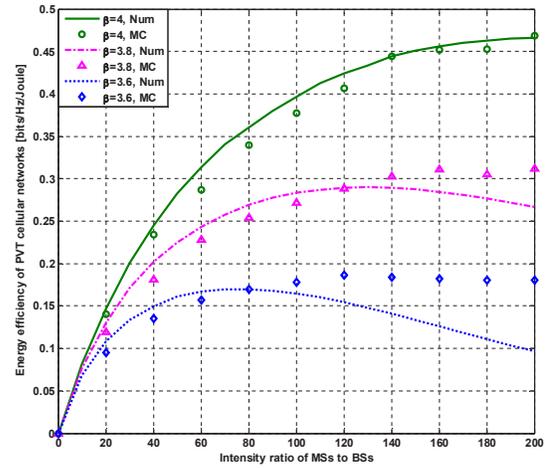}}
\caption{\small Energy efficiency of PVT cellular networks with respect to the intensity ratio of MSs to BSs considering the interfering link intensity.}
\label{fig10}
\end{figure}

In Fig.~\ref{fig10}, the effect of interfering link intensity on the energy efficiency of PVT cellular networks is investigated. When the BS intensity ${\lambda _B}$ is fixed, both numerical and MC simulation results show that the energy efficiency of PVT cellular networks is decreased when the interfering link intensity ${\lambda _{{\rm{Inf}}}}$ is increased from $3.0 \times {10^{ - 7}}{{\rm{\:m}}^{{\rm{ - 2}}}}$ to $5.0 \times {10^{ - 7}}{{\rm{\:m}}^{{\rm{ - 2}}}}$. Moreover, the maximal values of the energy efficiency in three different cases are 0.39, 0.29, and 0.23\:bits/Hz/Joule, which correspond to the intensity ratios of MSs to BSs as 170, 130, and 100, respectively. The MC simulation curves exhibit a good match with the numerical curves before the energy efficiency reaches the maximum, with deviations after the maximum value similar to the results shown in Fig. 9. These results imply that interference has obvious impact on the energy efficiency of PVT cellular networks.

\begin{figure}
\centerline{\includegraphics[width=7.2cm,draft=false]{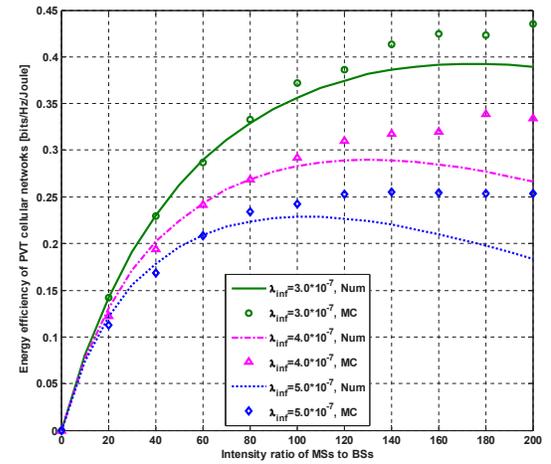}}
\caption{\small Energy efficiency of PVT cellular networks with respect to the intensity ratio of MSs to BSs considering the path loss exponent.}
\label{fig11}
\end{figure}

Finally, the impact of path loss exponent $\beta $ on the energy efficiency of PVT cellular networks is evaluated in Fig.~\ref{fig11}. Both numerical and MC simulation results illustrate that the energy efficiency of PVT cellular networks is increased when the path loss exponent is increased from 3.6 to 4. Moreover, the maximal values of the energy efficiency in three different cases are 0.17, 0.29, and 0.46\:bits/Hz/Joule, which correspond to the intensity ratios of MSs to BSs as 80, 130, and 190, respectively.

\section{Conclusions}
\label{sec6}
In this paper, we have proposed a novel model to evaluate the energy efficiency of PVT cellular networks considering spatial distributions of traffic load and power consumption. To derive this model, an aggregate traffic load model in a typical PVT cell is derived through its characteristic function based on stochastic cellular geometry. Moreover, taking into account path loss, log-normal shadowing and Rayleigh fading effects in wireless channels, an analytical total BS power consumption model in a typical PVT cell is derived. Based on the proposed energy efficiency model of PVT cellular networks, numerical results have shown that there is a maximal limit of energy efficiency in PVT cellular networks considering the tradeoff between the traffic load and BS power consumption. Moreover, wireless channel conditions have great impact on the energy efficiency of PVT cellular networks. Our analysis indicates that interference reduction or interference coordination can effectively improve the energy efficiency of PVT cellular networks, especially in scenarios with high intensity ratio of MSs to BSs. Furthermore, our results provide guidelines for the design of energy-efficient traffic control techniques and insights into the energy efficiency optimization in PVT cellular networks.

%
\appendix[Derivations of (\ref{eq25a}), (\ref{eq25b}) and (\ref{eq25c})]

Based on (\ref{eq23a}), we set ${R_1} = \left\| {{y_{Ik}} - {x_{Ik}}} \right\|$ in the following derivations. By conditioning on the BS Poisson point process ${\Pi _B}$ \cite{rf9}, the event of $\{ {R_1} > r|{\Pi _B}\} $ is equivalent to the event that no BSs  (${\Pi _B}$ points) locate inside the circle centered at ${x_{Ik}}$ with radius $r$, Then the conditional PDF of $R{}_1$ is given by
\begin{equation}
{f_{{R_1}|{\Pi _B}}}(r) =  - \frac{{d\Pr \{ {R_1} > r\} }}{{dr}} = 2\pi {\lambda _B}x{e^{ - \pi {\lambda _B}{r^2}}}.
\label{eq42}
\tag{29}
\end{equation}
Here the BS intensity ${\lambda _B}$ is used instead of the interfering link intensity ${\lambda _{\rm{Inf}}}$ due to the closest association rule under the stochastic geometry assumed previously.

The random variable ${R_2} = \left\| {{y_{Ik}} - {x_{M0}}} \right\|$ in (\ref{eq23a}) distributes in the same way as $\left\| {{y_{Ik}} - {y_{B0}}} \right\| (k = 1,2,3, \cdots )$ almost surely due to the stationarity of Poisson point processes, which further coincides with the distribution of $\left\| {{y_{Ik}}} \right\|(k = 1,2,3, \cdots )$ based on Slivnyak's theorem \cite{rf22}, behaving as if ${BS_{0}}$ locates at origin. Moreover, we set $\Psi = {S_k} \cdot {Q_k}$ for simplifying notation, where ${S_k}$ and $Q_k$ are independent in (\ref{eq23a}). Therefore, (\ref{eq23a}) is reduced as follows
\begin{equation}
{I_{{\rm{agg}}}} = \sum\limits_{k \in {\Phi _{{\rm{BS}}}}} {\frac{{R_1^\beta }}{{{{\left\| {{y_{Ik}}} \right\|}^\beta }}} \cdot {\Psi} \cdot {{\mathbf{1}}_{D1}}}  \cdot {{\mathbf{1}}_{D2}},
\label{eq43}
\tag{30}
\end{equation}
which is a marked Poisson point process defined on the BS Poisson point process ${\Pi _B}$. The intensity measure of (\ref{eq43}) is given by
\begin{equation}
\Lambda (dx,dr,d{\psi}) = {\lambda _{\rm{Inf}}}dx \cdot {f_{{R_1}| {\Pi _B}}}(r)dr \cdot {f_{\Psi}}(\psi)d\psi,
\label{eq44}
\tag{31}
\end{equation}
where ${\lambda _{\rm{Inf}}}$ is the interfering link intensity.

Based on (\ref{eq43}), (\ref{eq44}) and the Campbell theorem for marked Poisson point processes \cite{rf22}, a log-characteristic function of ${I_{{\text{agg}}}}$ is derived as follows
\begin{equation}
\begin{split}
 \ln {\phi _{{I_{\rm{agg}}}}}(\omega ) & = - {\left( {2\pi } \right)^2}{\lambda _{\rm{Inf}}}{\lambda _B} \\
 &\times{\int _0^{ \infty } \!\!\!\! {\underbrace {\int _0^{\infty } {\left[ {1 - {\phi _{\Psi}}\left( {{\frac {\omega {x^\beta }} {{y^\beta }}}} \right)} \right] ydy} }_{ = H\left(\omega ,x\right)} \cdot x{e^{ - 2\pi {\lambda _B}{x^2}}}dx}}  .
\end{split}
\label{eq45}
\tag{32}
\end{equation}
(\ref{eq45}) holds with ${\mathbf{E}}\left( {{{\mathbf{1}}_{D2}}\left\{ {{x_{Ik}} \in {\mathcal{C}_k}} \right\}} \right) = {e^{ - \pi {\lambda _B}{{\left\| {{y_{Ik}} - {x_{Ik}}} \right\|}^2}}}$ due to the fact that a point at ${x_{Ik}}$ belongs to a cell ${\mathcal{C}_k}$ if and only if there is no BS point locating inside the circle centered at ${x_{Ik}}$ with radius $\left\| {{y_{Ik}} - {x_{Ik}}} \right\|$.

Let $t = |\omega |{x^\beta }{y^{ - \beta }}$. Then $H(\omega ,x)$ in (\ref{eq45}) can be calculated by \cite{rf24}
\begin{equation}
\begin{split}
 H(\omega ,x) &= \frac{{\Gamma (2 - {\textstyle{2 \over \beta }})\cos ({\textstyle{\pi  \over \beta }})}}{{2 - {\textstyle{4 \over \beta }}}}|\omega {|^{\frac 2 \beta}}{x^2}{\mathbf{E}}({{\Psi}^{\frac 2 \beta }})\\
 & \quad \qquad \times \left[ {1 - j {\rm{sign}}(\omega ) \tan \left( {\frac{\pi }{\beta }} \right)} \right] .
\end{split}
\label{eq46}
\tag{33}
\end{equation}
Due to independence between ${S_k}$ and $Q_k$, we have
${\mathbf{E}}({{\Psi}^{2/\beta }}) = {\mathbf{E}}(S_k^{2/\beta }){\mathbf{E}}({Q_k^{2/\beta }})$.
%
Then by applying the integral
$\int_0^\infty  {{x^3}{e^{ - 2\pi {\lambda _B}{x^2}}}dy}  = {1 \mathord{\left/ {\vphantom {1 {8{{(\pi {\lambda _B})}^2}}}} \right.
 \kern-\nulldelimiterspace} {8{{(\pi {\lambda _B})}^2}}}$ \cite{rf25},
the characteristic function of ${I_{{\text{agg}}}}$ is derived as follows
\begin{equation}
{\phi _{{I_{{\text{agg}}}}}}(\omega ) = \exp \left\{ { - \delta |\omega {|^{{\frac 2 \beta }}}\left[ {1 - j {\text{sign}}(\omega )\tan \left( {\tfrac{\pi }
{\beta }} \right)} \right]} \right\},
\label{eq49a}
\tag{34a}
\end{equation}
with
\begin{equation}
\delta  = \frac{{{\lambda _{{\text{Inf}}}}}}
{{4{\lambda _B}}}\Gamma (1 - \tfrac{2}
{\beta })\cos (\tfrac{\pi }
{\beta }){\mathbf{E}}(S_k^{2/\beta }){\mathbf{E}}(Q_k^{2/\beta }),
\label{eq49b}
\tag{34b}
\end{equation}
\begin{equation}
{\mathbf{E}}({Q_k^{2/\beta }})  = \frac{{2\pi }}
{{\beta \sin (2\pi /\beta )}}\exp \left( {\tfrac{{4{c^2}{\sigma ^2}}}
{{{\beta ^2}}}} \right) .
\label{eq49c}
\tag{34c}
\end{equation}
This completes the derivations.


\end{document}